\documentclass[aip,jcp]{revtex4-1}

\usepackage[dvipsnames]{xcolor}
\usepackage{graphicx}

\draft

\begin{document}

\title{Collective and non-collective molecular dynamics in a ferroelectric nematic liquid crystal studied by broadband dielectric spectroscopy} 

\author{Aitor Erkoreka}
\affiliation{Department of Physics, Faculty of Science and Technology, University of the Basque Country UPV/EHU, Bilbao, Spain}
\author{Alenka Mertelj}
\affiliation{Jožef Stefan Institute, Ljubljana, Slovenia}
\author{Mingjun Huang}
\author{Satoshi Aya}
\affiliation{South China Advanced Institute for Soft Matter Science and Technology (AISMST), School of Molecular Science and Engineering, South China University of Technology, Guangzhou, China}
\affiliation{Guangdong Provincial Key Laboratory of Functional and Intelligent Hybrid Materials and Devices, South China University of Technology, Guangzhou, China}
\author{Nerea Sebastián}
\affiliation{Jožef Stefan Institute, Ljubljana, Slovenia}
\author{Josu Martinez-Perdiguero}
\email{jesus.martinez@ehu.eus}
\affiliation{Department of Physics, Faculty of Science and Technology, University of the Basque Country UPV/EHU, Bilbao, Spain}

\date{2023}

\begin{abstract}
A great deal of effort has been recently devoted to the study of dielectric relaxation processes in ferroelectric nematic liquid crystals, yet their interpretation remains unclear. In this work, we present the results of broadband dielectric spectroscopy experiments of a prototypical ferroelectric nematogen in the frequency range 10 Hz–110 MHz at different electrode separations and under the application of DC bias fields. The results evidence a complex behavior in all phases due to the magnitude of polar correlations in these systems. The observed modes have been assigned to different relaxation mechanisms based on existing theoretical frameworks. 
\end{abstract}


\maketitle

\section{Introduction}
Ever since the existence of ferroelectric liquid crystals (LCs) was predicted and confirmed by Meyer et al. in the 1970s,\cite{meyer_ferroelectric_1975} it was long thought that polar order and high fluidity were mutually exclusive properties.\cite{terzis_quantitative_1997, takezoe_antiferroelectric_2010}  The nomenclature of ferroelectric LCs was historically used to designate tilted smectic phases formed by chiral mesogens (SmC*), in which molecules are arranged in layers with their long molecular axis parallel to one another, and tilted at an angle with respect to the layer normal. Meyer el al. showed that if molecules possess a transverse component of the dipole moment, symmetry allows spontaneous polarization in the direction perpendicular to the tilt plane due to the chiral center.\cite{meyer_ferroelectric_1975} Due to the molecular chirality, the tilt plane additionally turns around the layer normal from layer to layer, and with it, the direction of polarization, resulting in a helical superstructure. The unwinding of this helix results in a macroscopic polarization with typical values of the order of nC$/$cm$^2$.\cite{terzis_quantitative_1997, kremer_broadband_2003} However, the recent discovery of the ferroelectric nematic (N$_{\mathrm{F}}$) phase has completely changed this notion, \cite{chen_first-principles_2020, sebastian_ferroelectric_2022} with polar ordering appearing in mesophases with no positional ordering. 

The conventional nematic (N) mesophase, in which rod-like molecules only possess orientational order, is nonpolar with as many dipoles pointing “up” as “down”, i.e. N phases possess inversion symmetry. The direction of preferred molecular orientation is denoted as the director $\mathbf{n}$, and then $\mathbf{n}$ and $-\mathbf{n}$ are equivalent. In the N$_{\mathrm{F}}$ phase there is a preferential direction for the dipoles, which breaks the inversion symmetry and gives rise to very large polarization values ($\mu$C$/$cm$^2$).\cite{nishikawa_fluid_2017, chen_first-principles_2020, chen_ideal_2022, brown_multiple_2021} In some compounds, another new intermediate phase between the N and N$_{\mathrm{F}}$ phases has been found, which has been described as an antiferroelectric splay nematic phase. \cite{mertelj_splay_2018, PhysRevLett.124.037801, sebastian_ferroelectric_2022, chen_smectic_2023} Still subject of debate, this phase has been referred to with different denominations such as M2,\cite{nishikawa_fluid_2017} N$_{\mathrm{X}}$, \cite{brown_multiple_2021} SmZ$_{\mathrm{A}}$ \cite{chen_smectic_2023} and N$_{\mathrm{S}}$ (splay nematic) \cite{mertelj_splay_2018} (throughout this work we will use the latter). Due to the novelty of these phases, research on this area has attracted considerable attention and their complete understanding is considered of fundamental importance in condensed matter physics and chemistry. Moreover, the properties of N$_{\mathrm{F}}$ materials have a wide array of potential high-impact applications such as in a new generation of LC displays, light modulators, memory devices, energy storage, etc.

Among the interesting properties of the N$_{\mathrm{F}}$ phase are the extremely large values of the dielectric permittivity (between $10^3$ and $10^5$) that have been reported.\cite{Nishikawa_giant, manabe_ferroelectric_2021, yadav_polar_2022, erkoreka_dielectric_2023} While Vaupotič et al. argue that these measurements reflect the actual material properties,\cite{vaupotic_dielectric_2023} Clark et al. have expressed caution about interpreting these results.\cite{clark_dielectric_2022} In any case, their theoretical models predict a thickness-dependent dielectric response, which has already been experimentally observed and quantified in the prototypical N$_{\mathrm{F}}$ RM734 material.\cite{erkoreka_dielectric_2023} Broadband dielectric spectroscopy (BDS) is a powerful technique to shed light on the molecular dynamics of these systems. In classical ferroelectric smectics, BDS evidences two dielectrically active collective modes corresponding to the fluctuations in the orientation of the polarization vector (Goldstone mode, fluctuations of the tilt direction) and to fluctuations in the amplitude of the polarization (soft mode, fluctuations of the tilt angle).\cite{kremer_broadband_2003, gouda_dielectric_1991} The Goldstone mode shows a negligible temperature dependence while, on the contrary, the soft mode is characterized by a critical slowing down of the relaxation frequency and a diverging amplitude near the transition to the higher temperature non-tilted phase (SmA).\cite{kremer_broadband_2003, lagerwall_book, gouda_dielectric_1991}

A similar thickness effect to that recently reported for the N$_{\mathrm{F}}$ phase\cite{erkoreka_dielectric_2023} was already observed in some SmC$^*$ compounds, which was explained by deviations from the ideal helicoidal structure arising due to strong polar anchoring on the measurement cells.\cite{haase_relaxation_2003}. However, in ferroelectric nematics, it is assumed that both the polarization and director are spatially uniform and parallel to the cell plates. As long as a thickness dependence of unknown origin is observed in the spectra, any investigation on the relaxation mechanisms of these materials must proceed with caution. In particular, the effect of DC bias fields could offer new insights into these issues. In this work, we present the results of a set of systematic BDS experiments in the N, N$_{\mathrm{S}}$ and N$_{\mathrm{F}}$ phases of the prototypical material DIO at different sample thicknesses and DC bias fields. BDS studies in N$_{\mathrm{F}}$ materials are often circumscribed to frequencies below $10$ MHz, which not only limits the range of relaxation processes that can be observed but can also hinder the interpretation of the lower frequency ones. Thus, an extended characterization up to $110$ MHz has been carried out. On the other hand, the results of the thickness-dependent dielectric response have been analyzed in the N, N$_{\mathrm{S}}$ and N$_{\mathrm{F}}$ phases. Additionally, we have extended the previously reported analysis of the effect of DC bias fields in the N$_{\mathrm{S}}$ phase \cite{brown_multiple_2021} to the N$_{\mathrm{F}}$ phase, thereby addressing the effect of a DC bias applied while cooling from the antiferroelectric N$_{\mathrm{S}}$ phase on the molecular dynamics of the system.

\section{Materials and Methods}

\subsection{Material}
\begin{figure}
\includegraphics[width=0.7\textwidth]{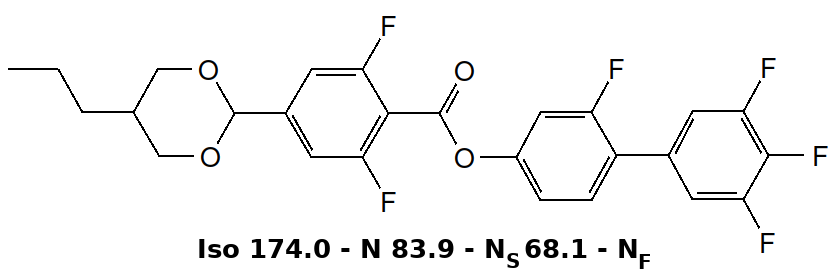}
\caption{\label{fig:DIOmolec} Chemical structure of DIO along with its transition temperatures.}
\end{figure}

The studied material is the ferroelectric nematogen DIO (2,3$^{\prime}$,4$^{\prime}$,5$^{\prime}$-tetrafluoro-[1,1$^{\prime}$-biphenyl]-4-yl 2,6-difluoro-4-(5-propyl-1,3-dioxan-2-yl) benzoate), which was synthesized via literature methods. Its molecular structure, phase sequence and transition temperatures are shown in Fig. \ref{fig:DIOmolec}. DIO molecules have a large dipole moment of $10.3$ D directed $14^{\circ}$ away from the long molecular axis.\cite{li_general_2022}

\subsection{Broadband dielectric spectroscopy}
The complex dielectric permittivity $\varepsilon(f) = \varepsilon'(f)-i\varepsilon''(f)$ was measured in the frequency range $f =$ 10 Hz–110 MHz. Low- and intermediate-frequency (up to 3 MHz) measurements were performed with an Alpha-A impedance analyzer from Novocontrol Technologies GmbH. The oscillator voltage was set to 0.03 V$_\mathrm{rms}$ for measurements without bias voltage. To obtain a better signal-to-noise ratio in the measurements under DC bias, the field was set to 0.01 V$_\mathrm{rms}/\mu\mathrm{m}$ , always well below the Fréedericksz threshold. The thickness dependence experiments were performed in custom-made gold-coated glass cells in a parallel plate configuration with no surface treatment because, although polyimide layers are useful to obtain a proper alignment, they act as an additional large capacitance in series with the LC cell, and can lead to other undesired effects such as charge accumulation.\cite{brown_multiple_2021, vaupotic_dielectric_2023, clark_dielectric_2022} Cells of different thicknesses were made using silica spheres with the final thickness determined by optical interferometry. The material was introduced into the cells by capillary action in the N phase at 115$^\circ$C to avoid sample degradation. The temperature during the measurements was controlled with a hot stage (Linkam). Qualitative polarizing optical microscopy (POM) observations with high intensity light were possible despite the high optical density of the gold layers. In order to perform  high-frequency measurements (up to 110 MHz), a dedicated setup comprising an HP 4294A impedance analyzer was employed. In this case, circular gold-plated brass electrodes were used, adjusting the electrode separation with rod-shaped silica spacers. The sample was then placed at the end of a modified HP 16091A coaxial test fixture, and a Quatro Cryostat from Novocontrol was employed for temperature control. The sample was first heated up to 115$^\circ$C at 5$^\circ$C/min to the N phase and stabilized for 5 minutes at this temperature. Measurements were then carried out on cooling at 0.25$^\circ$C/min. The complex dielectric permittivity was determined by dividing the measured capacitance by the capacitance of the empty cells. The stray capacitance of the measurement circuit was carefully taken into account in all cases. For a quantitative analysis of the dielectric relaxation processes, the data were fitted to the Havriliak-Negami (HN) formula with a conductivity term:

\begin{equation}
    \varepsilon (f) = \sum_{k} \frac{\Delta \varepsilon_k}{\left[1+\left(i \frac{f}{f_k}\right)^{\alpha_k} \right]^{\beta_k}} + \varepsilon_{\infty} +\frac{\sigma}{\varepsilon_0(i\,2\pi f)^{\lambda}}\mathrm{,}\label{HN_eq}
\end{equation}

\noindent
where $\Delta\varepsilon_k$, $f_k$, $\alpha_k$ and $\beta_k$ are respectively the dielectric strength, relaxation frequency and broadness exponents of mode $k$, $\varepsilon_{\infty}$ is the high-frequency dielectric permittivity, $\sigma$ is a measure of the conductivity, and $\lambda$ is an exponent between 0 and 1.

\section{Results and discussion}

\begin{figure}
\includegraphics[width=0.6\textwidth]{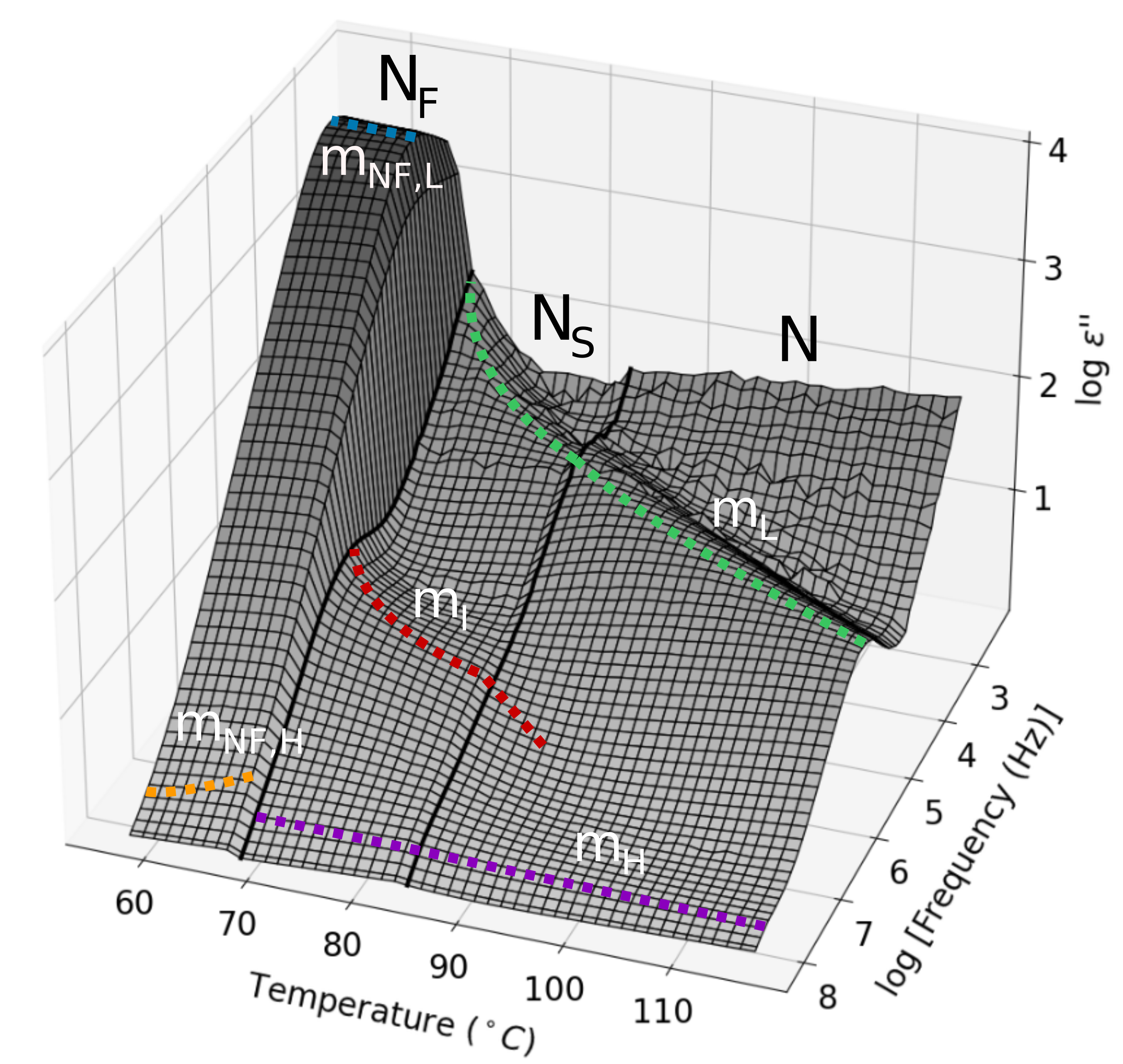}
\caption{\label{fig:4294-3D} Three-dimensional plot of the imaginary component of the complex dielectric permittivity as a function of frequency and temperature for a 40 $\mu$m thick cell. The dark black lines indicate the transition temperatures of the various phases. The dashed lines are guides for the eyes and color coded according to Fig. \ref{fig:4294-fit}.}
\end{figure}

First, the main features of the dielectric spectra of DIO from 1 kHz up to 110 MHz will be discussed. This frequency range allows the identification of all the relevant dielectrically active modes of this material.  Fig. \ref{fig:4294-3D} shows the imaginary component of the complex dielectric permittivity of DIO as measured on cooling in a $d=40\;\mu$m thick cell. In the N phase, we can discern two dielectrically active modes, referred to in this work as m$_\mathrm{L}$ and m$_\mathrm{H}$, where L and H stand for low- and high-frequency respectively. At 84$^{\circ}$C, the N$_{\mathrm{S}}$ phase is reached, marked by a small discontinuity in $\varepsilon''$. In the N$_{\mathrm{S}}$ phase, m$_\mathrm{L}$ and m$_\mathrm{H}$ are still present, but a new intermediate-frequency mode appears, denoted by m$_\mathrm{I}$. Lastly, the N$_{\mathrm{F}}$ phase is reached at 68$^\circ$C, characterized by a very large increase in $\varepsilon''$ where another two relaxation processes are observed: m$_\mathrm{NF,L}$ being the most prominent one and m$_\mathrm{NF,H}$ at higher frequencies. The sample then crystallizes at 56$^\circ$C.

\begin{figure}
\includegraphics[width=0.5\textwidth]{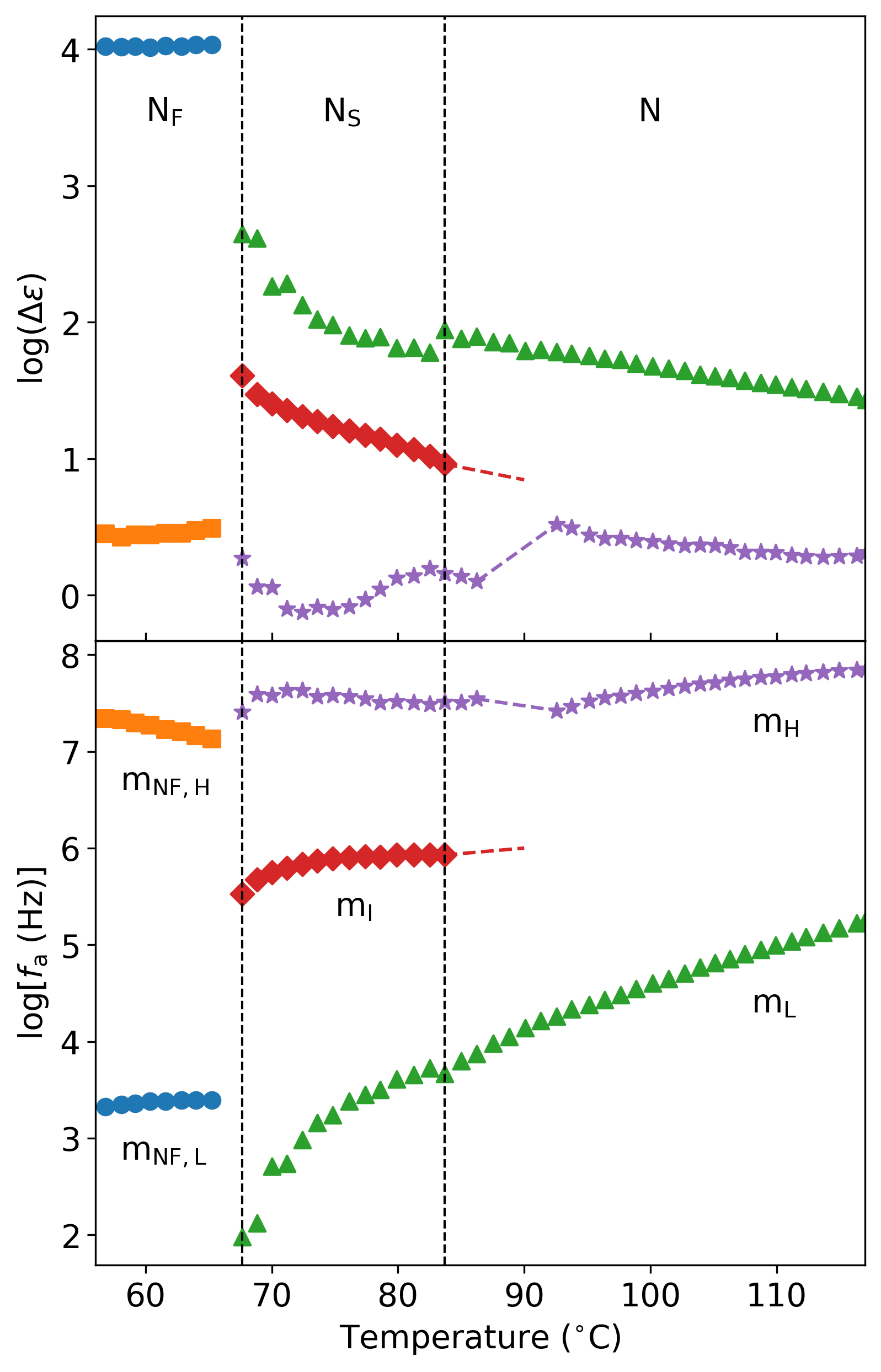}
\caption{\label{fig:4294-fit} Dielectric strengths ($\Delta\varepsilon$) and frequencies of maximum absorption ($f_a$) of the observed modes in a 40 $\mu$m thick cell as a function of temperature obtained from fits to the HN formula (Equation \ref{HN_eq}).}
\end{figure}

In order to study the aforementioned relaxation processes more carefully, the experimental data were fitted to HN relaxations (see Equation \ref{HN_eq}). The obtained dielectric strengths and frequencies of maximum absorption are presented in Fig. \ref{fig:4294-fit} (fit examples and parameters can be found in Fig. S1 and Table S1). These results agree with those reported by Brown et al.,\cite{brown_multiple_2021} with the addition that we are able to observe further high-frequency processes. It is worth noticing that the permittivity values in both the N and N$_{\mathrm{S}}$ phases are quite large if compared to regular nematics, which can be expected in highly polar molecules such as DIO. Regarding the molecular alignment, the cell for high-frequency measurements does not allow for direct visual inspection. However, because the measured permittivity value is one order of magnitude larger than the one obtained in cells for which planar alignment can be deduced from POM (as will be shown below), we concluded that this cell favors homeotropic-like alignment in the N phase. Such homeotropic-like alignment in this kind of cells is consistent with the strong tendency towards out-of-plane alignment already observed for RM734 in the same cells. \cite{erkoreka_dielectric_2023} According to the Nordio-Rigatti-Segre theory,\cite{kremer_broadband_2003, luigi_nordio_dielectric_1973} m$_\mathrm{L}$ can be associated with the rotation of individual molecules around their short axis, i.e. fluctuations of the longitudinal component of the dipole moment $\mu_\mathrm{L}$ around the short axis of the mesogenic unit, and is also known as $\delta$-process. The absorption frequency of this mode follows the Arrhenius equation $f_a = f_\infty \exp\left(-E_{\mathrm{A}}/k_{\mathrm{B}}T\right)$, where $f_\infty$ is a prefactor, $E_{\mathrm{A}}$ is the activation energy, $k_{\mathrm{B}}$ is the Boltzmann constant and $T$ is the temperature. A fit to this function yields an activation energy of 110 kJ/mol (see Fig. S2). This is a high activation energy, which is expected for the $\delta$-process since the nematic potential hinders this kind of rotation. As a comparison, the activation energy of this mode in 7CB, a conventional nematic LC, is $91.1$ kJ/mol.\cite{kremer_broadband_2003} However, it is important to point out that m$_\mathrm{L}$ involves at least some degree of collective behavior. This is evident from the large amplitude of this process, which can not be uniquely explained by the large dipole moment of DIO molecules. Furthermore, the polar correlations become even more pronounced upon the transition to the antiferroelectric N$_{\mathrm{S}}$ phase, where this process exhibits soft mode behavior. This mode, m$_\mathrm{L}$, corresponds to the low-frequency process observed in RM734 in equivalent measurements, see Fig. S3.\cite{erkoreka_dielectric_2023, PhysRevLett.124.037801} In RM734, with direct N-N$_{\mathrm{F}}$ transition, m$_\mathrm{L}$ exhibits strong softening already in the N phase, for which it has been shown that the growth of polar order is coupled to splay fluctuations.\cite{PhysRevLett.124.037801} In the case of DIO, it has been proposed that, in the N$_{\mathrm{S}}$ phase,  the periodicity at which polarization changes sign is 180 \AA{}.\cite{chen_smectic_2023} Moreover, Emelyanenko et al. have theoretically modeled all three phases of DIO, and they have suggested that the size of these polar domains increases with increasing temperature in the N$_{\mathrm{S}}$ phase and diverges at the N$_{\mathrm{S}}$-N transition.\cite{emelyanenko_emergence_2022} Consequently, it is plausible that short-range correlations between dipoles start to occur already in the N phase accounting for the relatively large $\varepsilon'$ observed already in the non-polar phase. This has recently been proposed from optical and electrooptical experiments as well.\cite{yadav_spontaneous_2023} The mode m$_\mathrm{H}$, on the contrary, can be attributed to the rotation of individual molecules around their long axis (called $\beta$-process), which involves fluctuations in the transverse component of the dipole moment $\mu_T$ projected along the direction of the probe electric field. It is also a temperature-activated process, with an activation energy of 36 kJ/mol (see Fig. S2). This lower activation energy indicates that this mode is not as influenced by the molecular environment as the $\delta$-process. In the N$_{\mathrm{S}}$ phase, m$_\mathrm{I}$ also exhibits soft mode behavior near the transition to the N$_{\mathrm{F}}$ phase, which suggests it might be related to some kind of collective fluctuation. It should be noted that near the N-N$_{\mathrm{S}}$ transition m$_\mathrm{I}$ begins to be discernible even in the N phase, however, its accurate deconvolution is not possible. For completeness, the dashed line in Fig. \ref{fig:4294-fit} marks the range of temperatures at which the mode is discerned.


\begin{figure}
\includegraphics[width=0.9\textwidth]{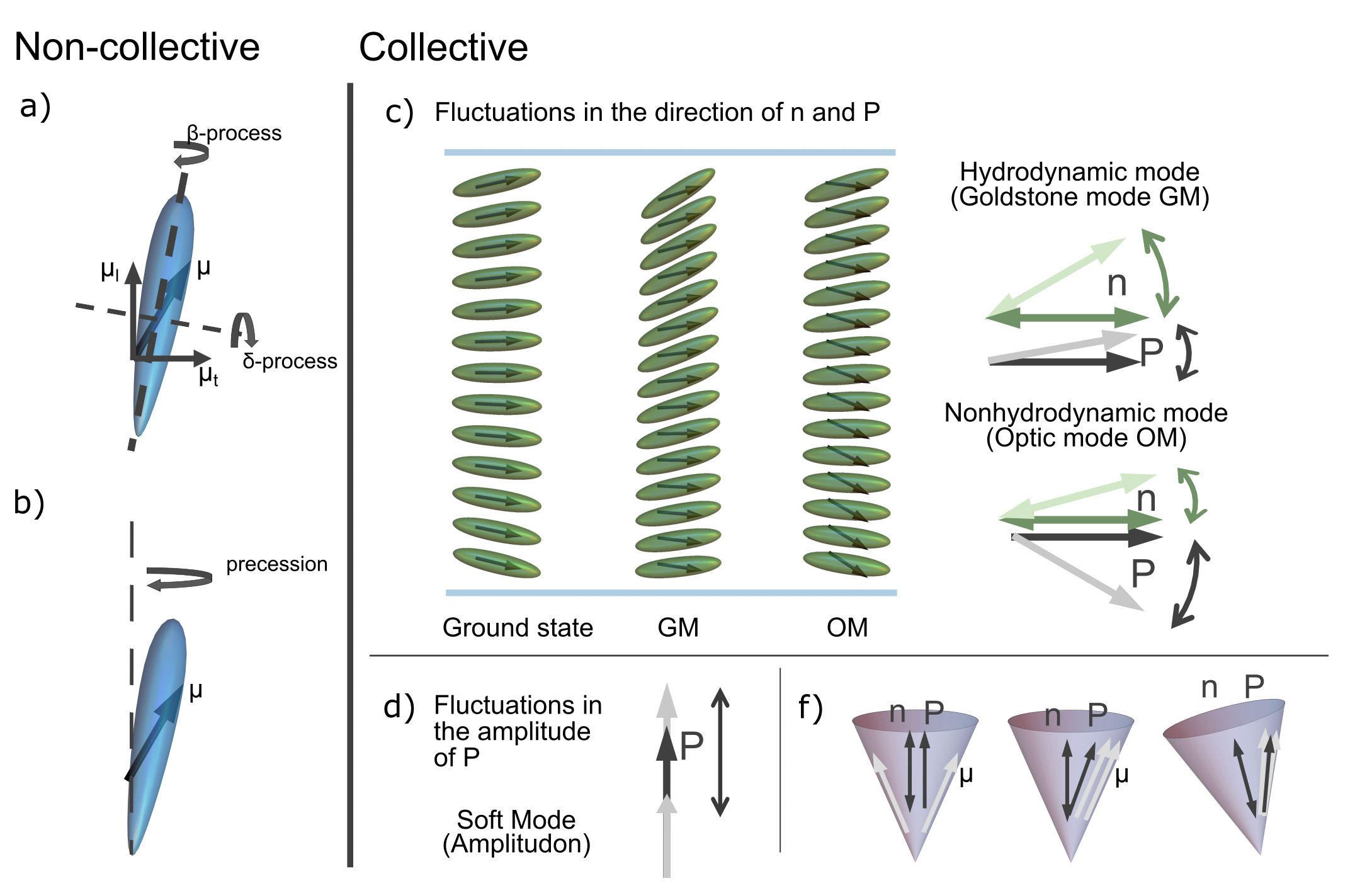}
\caption{\label{fig:sketch} Schematic representation of the proposed non-collective and collective modes. Non-collective molecular rotational modes contributing to dielectric relaxations can be described as (a) molecular rotation around the short ($\delta$-process) and long ($\beta$-process) molecular axis, and (b) the precessional motion about the director. (c) Collective fluctuations in the direction of the nematic director $\mathbf{n}$ and the polarization $\mathbf{P}$. Hydrodynamic and optic (non-hydrodynamic) modes for the fluctuations in the direction of $\mathbf{n}$ and $\mathbf{P}$. Representation for thin cells, where green ellipsoid accounts for the local director orientation and black arrows correspond to the local polarization direction, in correspondence with the theoretical model proposed in the work by Vaupotič et al.\cite{vaupotic_dielectric_2023} Arrow representation in the right visualizes that the hydrodynamic GM mode is dominated by director fluctuations in phase with the smaller polarization fluctuations, while the higher frequency OM is dominated by fluctuations of polarization with $\mathbf{n}$ and $\mathbf{P}$ in counter phase. This is analogous to the hydrodynamic and optic modes reported in the case of suspensions of ferroelectric nanoparticles in nematic liquid crystals.\cite{copic_nano_2007, sebastian_comparison_2018} (d) Representation of the fluctuations in the amplitude of $\mathbf{P}$, i.e. soft mode (amplitudon). (f) Dipole moment of DIO is directed $14^{\circ}$ away from the long molecular axis.\cite{li_general_2022}  In the uniaxial state dipoles are randomly oriented as represented by the cone.\cite{mandle_molecular_2021} Collective precession and reorientation around the molecular short axis allow for illustration of the hydrodynamic and optic modes depicted in (c).}
\end{figure}

Finally, two modes appear upon the transition to the N$_{\mathrm{F}}$ phase. The most prominent one is m$_\mathrm{NF,L}$, which shows a negligible temperature dependence in both strength and relaxation frequency. This process could be the Goldstone mode that is expected to appear in the N$_{\mathrm{F}}$ phase. In the continuous phenomenological model (CPM) developed by Vaupotič et al., \cite{vaupotic_dielectric_2023} the minimization of the free energy for a system with coupled polarization and director fluctuations yields three dielectrically active collective modes: two so-called phason modes related to the fluctuations in the direction of the director and of the polarization, respectively, and one soft mode (SM) related to the fluctuations in the amplitude of the polarization (so-called amplitudon). In order to differentiate the two phason modes more clearly, we will refer to the slow process as Goldstone mode (GM, equation 12 in Ref. \onlinecite{vaupotic_dielectric_2023}) and to the fast process as optic mode (OM, equation 13 in Ref. \onlinecite{vaupotic_dielectric_2023}). While the director and polarization fluctuate in phase in the GM, they fluctuate in counter phase in the OM. Additionally, the GM (associated primarily with director orientation fluctuations) dominates the dielectric spectrum, while the authors expect the OM (dominated by fluctuations in the direction of polarization) to appear in the MHz regime. The SM, on the other hand, is characterized by a diverging amplitude and slowing down of the relaxation frequency near the transition to the ferroelectric phase. These theoretical results could be consistent with our data, since the dominant m$_\mathrm{NF,L}$ could be attributed to the GM and m$_\mathrm{NF,H}$ to the OM, being m$_\mathrm{L}$ the SM in the N$_{\mathrm{S}}$ phase. A similar interpretation was proposed by Yadav et al. for m$_\mathrm{L}$ (P1 in their paper).\cite{yadav_polar_2022}  However, we have to note that in our measurements the amplitude of m$_\mathrm{NF,H}$ is small and it is also compatible with the $\beta$-relaxation (or a mixture of both effects). A visual representation of the proposed relaxation mechanisms can be seen in Fig. \ref{fig:sketch}.

\begin{figure}
\includegraphics[width=0.6\textwidth]{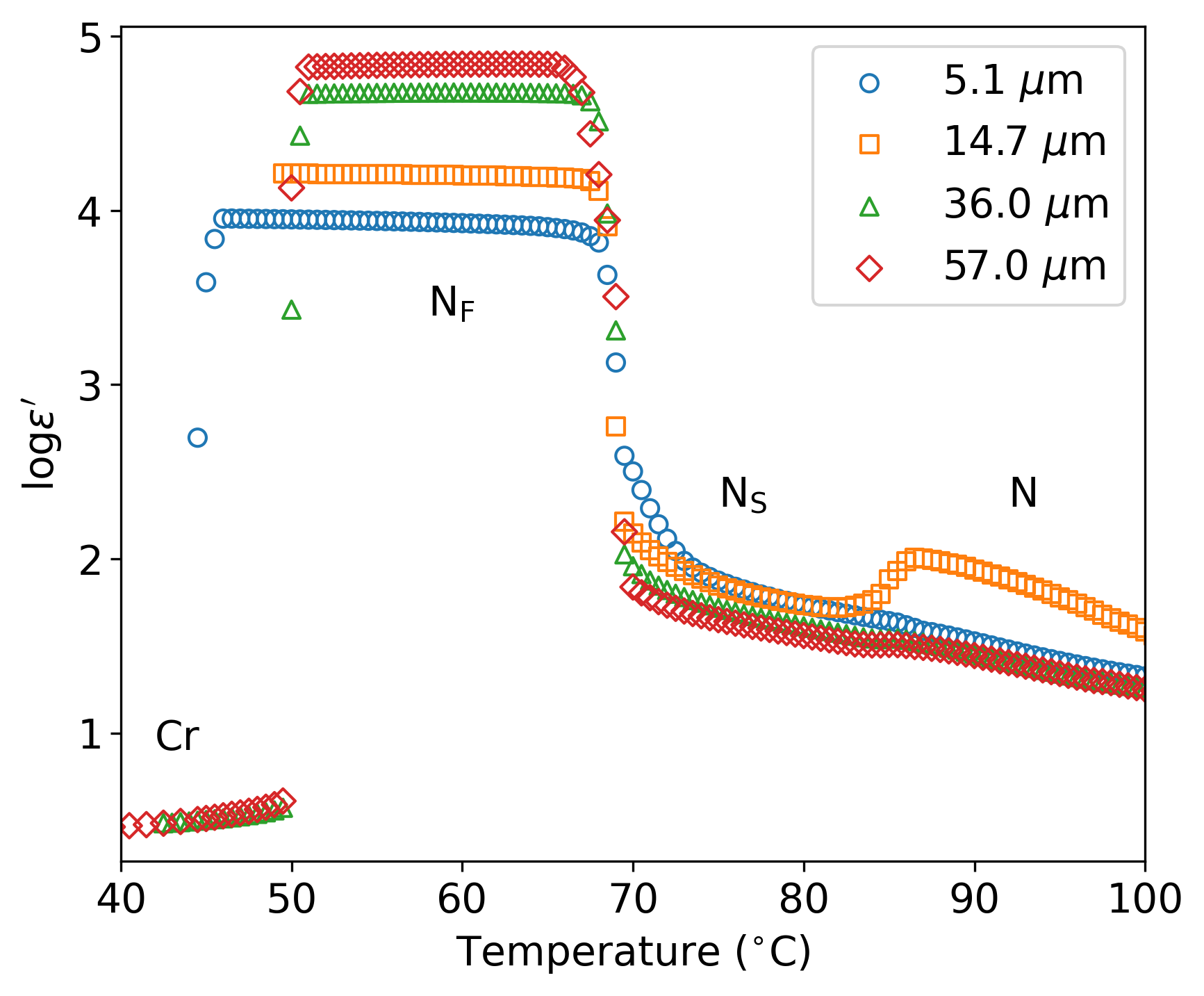}
\caption{\label{fig:all_ds-100Hz} Temperature dependence of the real component of the measured complex dielectric permittivity for various cell thicknesses at 100 Hz.}
\end{figure}

In order to systematically study the cell thickness effects in this material, we prepared four different glass cells with sputtered gold electrodes and thicknesses ranging from 5.1 to 57 $\mu$m. Fig. \ref{fig:all_ds-100Hz} shows the real component of the measured dielectric permittivity as a function of temperature at 100 Hz for all sample thicknesses. Starting from the high-temperature N phase, we observe that in this type of cells the alignment reached is not hometropic-like and birefringent domains are observed, suggesting a random planar orientation of the sample (see Fig. S4). The transition to the N$_{\mathrm{S}}$ ($\sim 86^{\circ}$C) phase is marked by a small bump in $\varepsilon'$, especially noticeable in the 14.7 $\mu$m thick cell, where the difference in value can be attributed to a slightly different alignment. On cooling, the transition to the N$_{\mathrm{F}}$ phase (69$^{\circ}$C) is marked by a very large increase of $\varepsilon'$. In this phase, a strong thickness dependence is readily observed, in which the thicker the measurement cell, the larger the permittivity. As a matter of fact, $\varepsilon'$ is increased by about an order of magnitude from 5.1 $\mu$m to 57 $\mu$m, pointing towards a linear dependence on the sample thickness $d$. In these runs, the N$_{\mathrm{F}}$ phase is stable within a temperature interval of $\sim 20^\circ$C except in the 5.1 $\mu$m thick cell, in which the sample crystallized at a slightly lower temperature. As expected, the dielectric response of the crystal (Cr) phase does not exhibit a thickness dependence, and the values of $\varepsilon'$ agree for all thicknesses.

\begin{figure*}
\includegraphics[width=\textwidth]{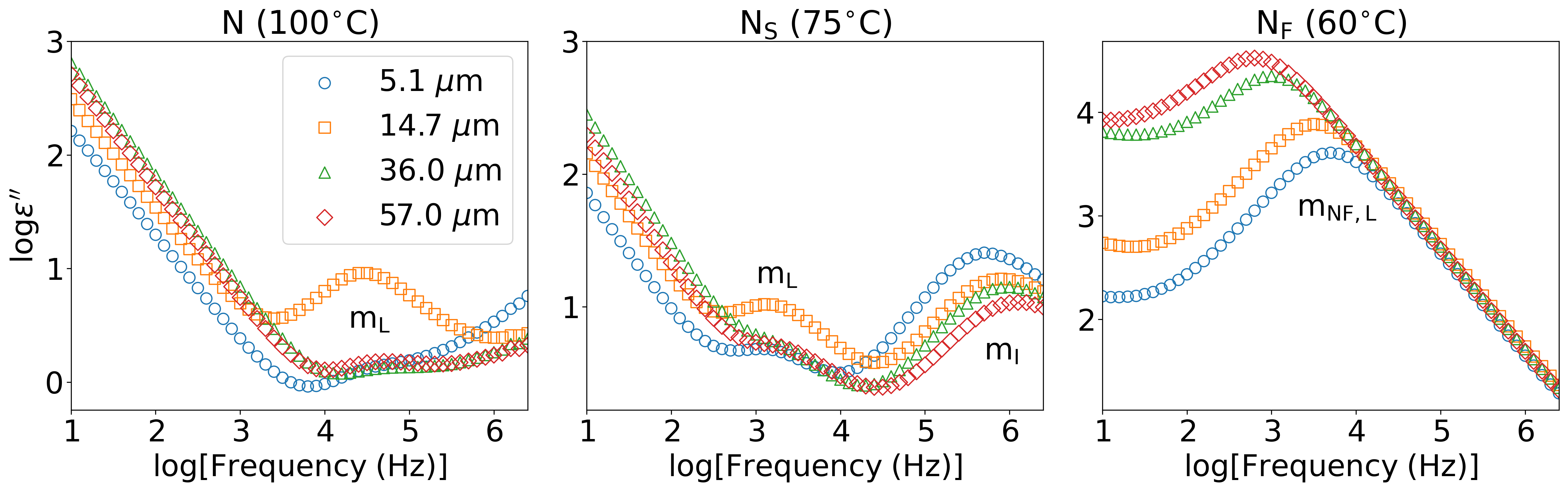}
\caption{\label{fig:N-NS-NF-spectra} Spectrum of the imaginary component of the dielectric permittivity measured in the N, N$_{\mathrm{S}}$ and N$_{\mathrm{F}}$ phases for various sample thicknesses.}
\end{figure*}

\begin{figure}
\includegraphics[width=0.5\textwidth]{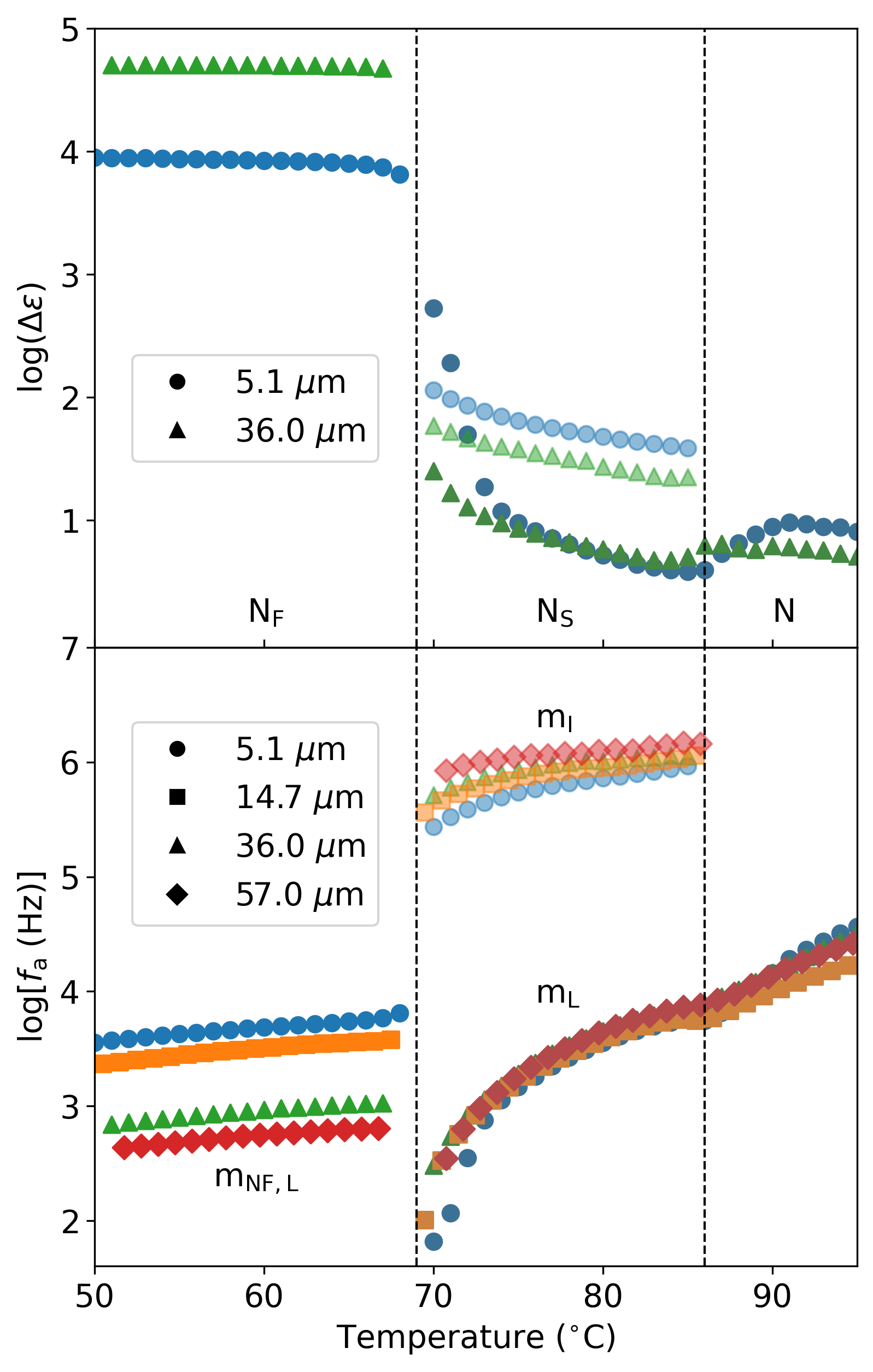}
\caption{\label{fig:ds_fits} Dielectric strengths ($\Delta\varepsilon$) and frequencies of maximum absorption ($f_a$) of the observed modes as a function of temperature obtained from fits to the HN formula (Equation \ref{HN_eq}) for various sample thicknesses. In the upper graph, only two thicknesses are shown for clarity.}
\end{figure}

The relaxation processes measured in these cells can be identified in the dielectric absorption spectra presented in Fig. \ref{fig:N-NS-NF-spectra} for the three LC phases, as well as in the temperature evolution of their frequencies and amplitudes shown in Fig. \ref{fig:ds_fits} for all thicknesses, as obtained from the fits to the HN formula (Equation \ref{HN_eq}) (only the amplitudes of two thicknesses are shown for clarity). Like in the results presented above with gold-plated brass electrodes, in the N phase, the m$_\mathrm{L}$ and m$_\mathrm{H}$ processes can be observed. The larger strength of m$_\mathrm{L}$ in the 14.7 $\mu$m thick cell is attributed to the alignment effect previously inferred from Fig. \ref{fig:all_ds-100Hz}. In fact, a greater relaxation amplitude is expected if the alignment is somewhat more homeotropic-like, since the fluctuations in $\mu_\mathrm{L}$ projected along the direction of the probe electric field would be greater in that case.  At higher frequencies the onset of m$_\mathrm{H}$ appears, although its relaxation frequency is far beyond the accessible frequency range in this type of cells. At low frequencies, another absorption process of great amplitude takes place, which appears as a straight line due to the limited frequency range, and is related to conductivity and electrode polarization effects (see also Fig. S5). In the N$_{\mathrm{S}}$ phase, we again observe m$_\mathrm{L}$ and m$_\mathrm{I}$. In the N and N$_{\mathrm{S}}$ phases, aside from slight differences in strength, the temperature evolution of the modes is equivalent. We observe a small but systematic thickness dependence in both the slope and value of the m$_\mathrm{I}$ frequencies, which might an issue worth investigating in the future.

\begin{figure}
\includegraphics[width=0.6\textwidth]{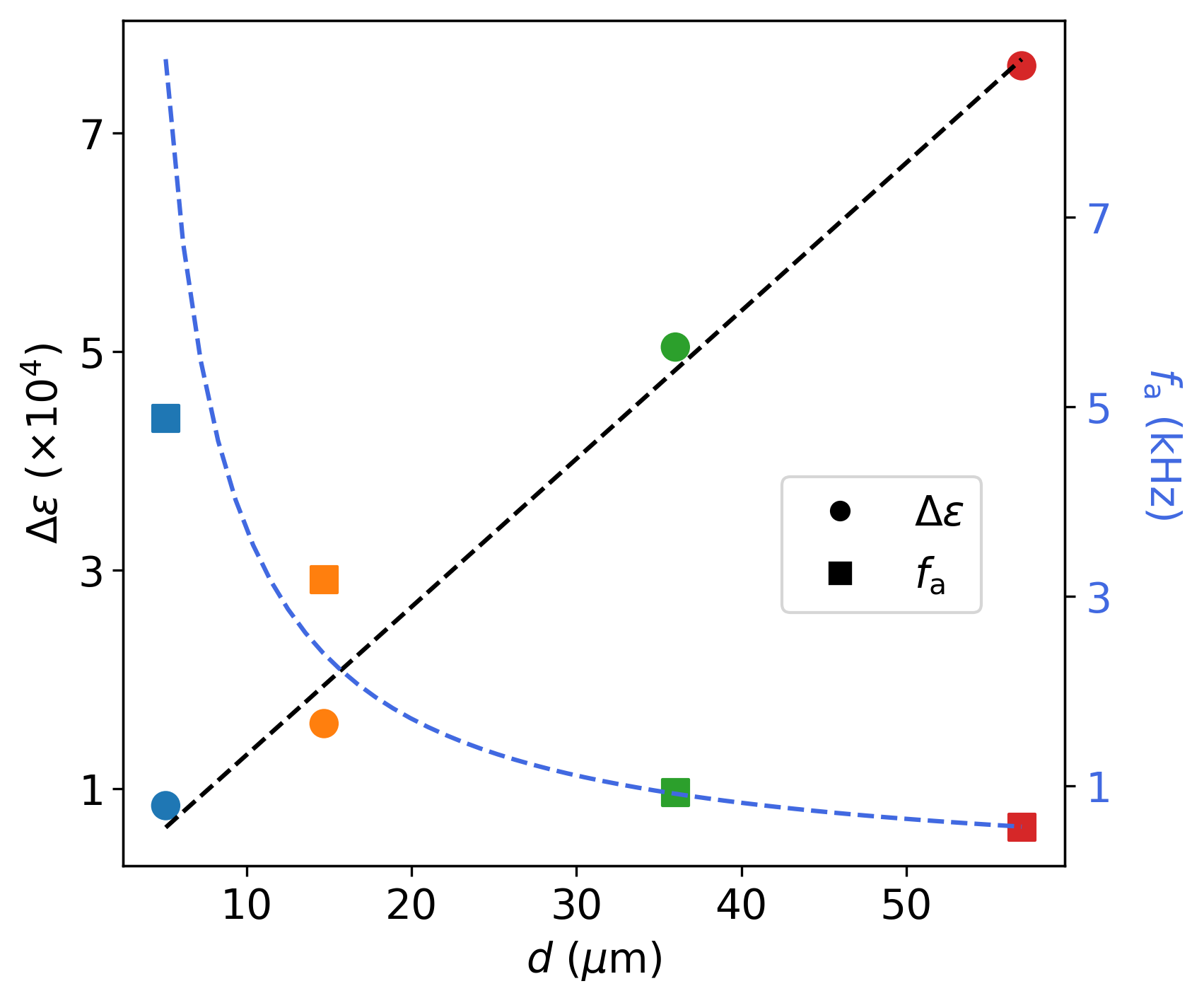}
\caption{\label{fig:PCG} Thickness dependence of the dielectric strength (circles) and frequency of maximum absorption (squares) of m$_\mathrm{NF,L}$ at 60$^\circ$C. The dashed lines are fits to the experimental data.}
\end{figure}

In the N$_{\mathrm{F}}$ phase, only  m$_\mathrm{NF,L}$ is seen in the spectra. In addition to the strong thickness effect on the relaxation strength already inferred from Fig. \ref{fig:all_ds-100Hz}, the dependence on the thickness of the relaxation frequency is also evident.
Such thickness effect is summarized in Fig. \ref{fig:PCG}, which shows the dielectric strength and frequency of maximum absorption of m$_\mathrm{NF,L}$ as a function of $d$. As can be inferred from Figs. \ref{fig:all_ds-100Hz}, \ref{fig:N-NS-NF-spectra} and \ref{fig:ds_fits}, the dielectric strength qualitatively shows a linear dependence on d, while the relaxation frequency is inversely proportional to $d$. This is the same behavior reported by our group for RM734.\cite{erkoreka_dielectric_2023} In that work, the results were analyzed on the basis of three different mechanisms. Clark et al. proposed that the GM that is expected to be present in ferroelectric nematics would be responsible for the large values of the dielectric permittivity and for the thickness effect.\cite{clark_dielectric_2022} The minimization of the free energy of the N$_{\mathrm{F}}$ system renders the polarization spatially uniform and parallel to the electrodes. The GM would then correspond to a reorientation of the polarization vector upon the application of an external electric field. This phenomenon, coupled with the high fluidity of these materials, would screen the electric field in the bulk via the charging of the interfacial layers that bound the sample in typical measurement cells, giving rise to a so-called polarization-external capacitance Goldstone reorientation mode (PCG). The authors argue that the measurement results should therefore be interpreted as apparent values, and they predict a Debye-type relaxation process, with $\Delta\varepsilon\propto d $ and $f_r\propto 1/d$,\cite{clark_dielectric_2022} in line with our results (see Fig. \ref{fig:PCG}).

On the other hand, the CPM model developed by Vaupotič et al. includes thickness dependent terms and predicts a relaxation frequency with a $1/d$ dependence (Equation 12 in Ref. \onlinecite{vaupotic_dielectric_2023}). Although they neglect them based on their qualitative observations, their inclusion would also be in line with our experimental results and would suggest that the dominant term in their model would be associated with flexoelectricity. Barthakur et al. recently measured a very large flexoelectric effect in RM734 which could explain this dominance.\cite{flexoelectric} Unfortunately, to the best of our knowledge, no measurements of flexoelectric coefficients have been performed for DIO, which could help setting a common ground for this characteristic effect of the N$_{\mathrm{F}}$ phase. It is also interesting to note that Emelyanenko et al. have theoretically studied the flexoelectric effect in DIO starting from an intermolecular interaction of a particular symmetry.\cite{emelyanenko_emergence_2022} In fact, they argue that it is an effect of great importance, since it appears to be the generator of both the splayed structure and the improper polarization present in the N$_{\mathrm{S}}$ phase. Returning to the CPM model, the lower relaxation frequency of thicker cells might be suggesting that the director has more freedom to reorient (less energy cost, lower frequency) than in thinner cells.

As discussed in a previous article,\cite{erkoreka_dielectric_2023} it is well known that the electrode polarization (EP) effect can create a thickness dependent dielectric response and give rise to extremely large permittivity values.\cite{european, chassagne_compensating_2016} This phenomenon arises due to the accumulation of ions on the electrode surface, and is present in all kinds of substances at low enough frequencies. These charged layers screen the electric field in the bulk, manifesting itself as a thickness dependent relaxation process that can potentially mask the intrinsic response of the material. Several works predict that an EP affected spectrum will show a permittivity value proportional to $d$ below a given frequency, the latter being inversely proportional to $d$.\cite{chassagne_compensating_2016, sawada_complex_1998, klein_modeling_2006, kondratenko_ionic_2018} This behavior is again compatible with the present results (see Fig. \ref{fig:PCG}). N$_{\mathrm{F}}$ materials such as DIO or RM734 can be expected to contain larger ion concentrations than conventional nematics\cite{basnet_soliton_2022, patterning} due to their large molecular polarity. However, one should note that the bulk electrical conductivity associated with the reorientation of the polarization vector in the N$_{\mathrm{F}}$ phase according to the PCG model is $\sigma=P^2/\gamma$, where $P$ is the magnitude of polarization and $\gamma$ is the rotational viscosity. Introducing $P=6$ $\mu$C$/$cm$^2$ and $\gamma=0.25$ Pa s,\cite{chen_ideal_2022} then $\sigma \sim 10^{-2}$ S/m. This high effective conductivity along with a polarization self-screening length $\xi_{P}$ in the nm range, much smaller than any achievable ionic screening length,\cite{clark_dielectric_2022} rules out the possibility that EP effects are behind the large permittivity values and the thickness effect.

Another interfacial effect similar to EP related to charge accumulation that can also lead to dielectric dispersion and absorption at low frequencies is the well-known Maxwell-Wagner-Sillars (MWS) effect. This phenomenon occurs due to the build-up of charges across internal interfaces in heterogeneous systems.\cite{kremer_broadband_2003} In this study, as explained earlier, we did not use any alignment layers in order to avoid undesired effects in the dielectric measurements. However, this makes both the director and polarization vector non-uniform throughout the sample, leading to the accumulation of bound charge within the material. Consequently, the MWS effect should also be taken into consideration.

\begin{figure}
\includegraphics[width=0.5\textwidth]{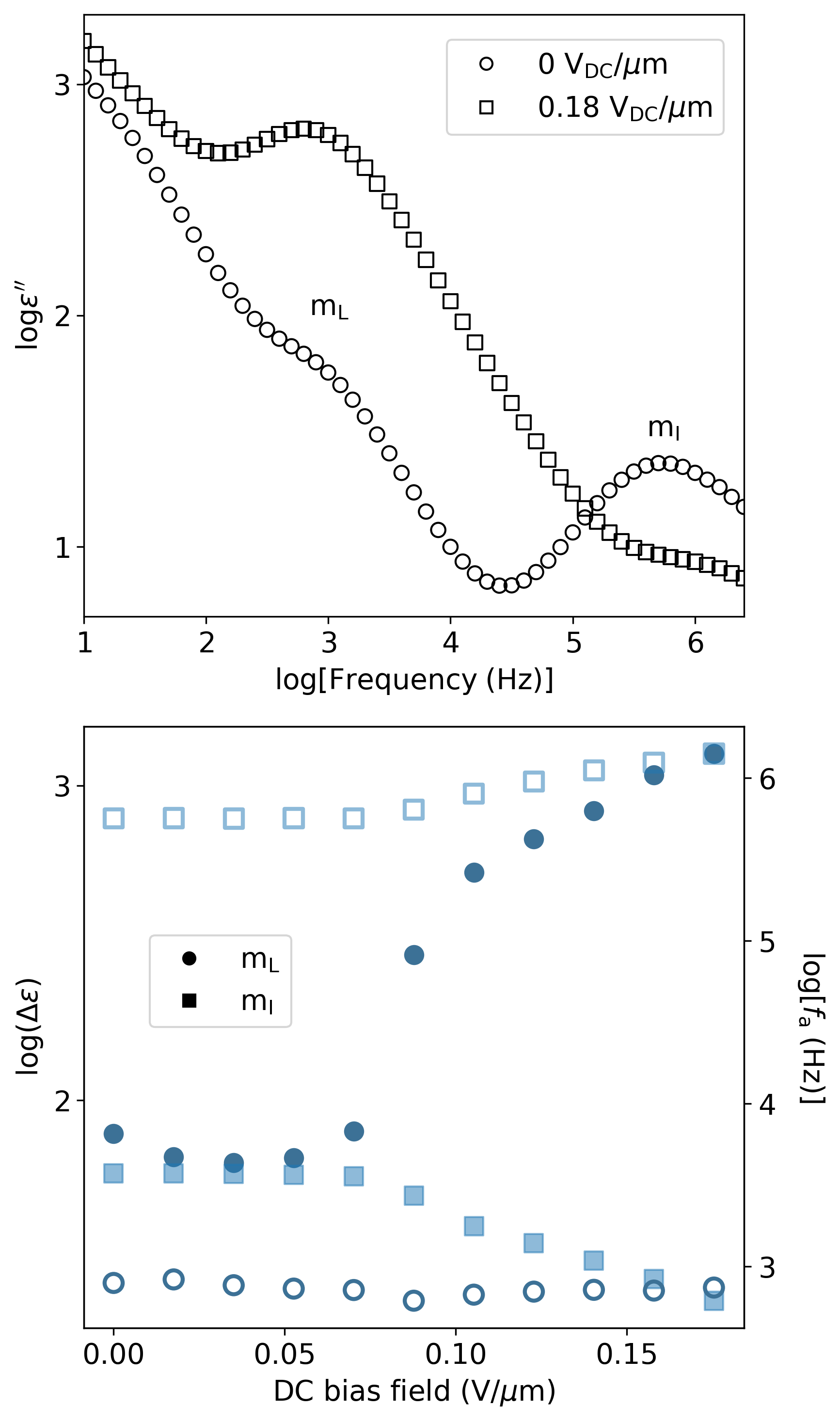}
\caption{\label{fig:fieldNS} Effect of a DC bias on the dielectric spectrum of the N$_{\mathrm{S}}$ phase at 76$^\circ$C in a 5.1 $\mu$m thick cell. Top: Dielectric loss spectrum at 0 V$_\mathrm{DC}$/$\mu$m and 0.18 V$_\mathrm{DC}$/$\mu$m. Bottom: Dielectric strengths ($\Delta\varepsilon$, solid symbols) and frequencies of maximum absorption ($f_\mathrm{a}$, empty symbols) of the observed modes as a function of the external DC bias field.}
\end{figure}

Additional insights into molecular mechanisms can be obtained from the dielectric characterization under DC bias fields. Experiments have been performed for different cell thicknesses. However, we limit this part of the study to thin cells ($\sim 5 \; \mu$m) due to the large electroconvective currents and sample degradation observed when large DC bias fields (up to 1 V$_\mathrm{DC}$/$\mu$m) were applied in thicker cells (see Fig. S6). After each measurement, the bias field was removed and the resulting spectrum was compared to the original one to ensure that the sample was not degraded. In the first set of experiments the cell was heated up to 110$^{\circ}$C and then cooled at 1$^{\circ}$C/min to the desired temperature for DC bias application. Fig. \ref{fig:fieldNS} shows the DC bias field dependence of the two modes of the N$_{\mathrm{S}}$ phase at 76$^{\circ}$C. On the one hand, it is observed that, while the strength of m$_\mathrm{L}$ increases, that of m$_\mathrm{I}$ decreases. Comparing this with the results of Fig. \ref{fig:4294-fit}, in which the strength of m$_\mathrm{L}$ is larger than that of m$_\mathrm{I}$, we conclude that we are certainly inducing a homeotropic-like alignment. Additionally, it would be consistent with the previous assignment of m$_\mathrm{L}$ to the SM, since a DC bias would increase the magnitude of the amplitude fluctuations.  On the other hand, the relaxation frequency of m$_\mathrm{L}$ stays practically constant, while that of m$_\mathrm{I}$ slightly increases. It is interesting to note, however, that these features only start to happen from a threshold field of around 0.08 V$_\mathrm{DC}$/$\mu$m. These results agree with those of Brown et al.\cite{brown_multiple_2021} It is also worth noticing that Nishikawa et al. reported, for fields above a threshold of 0.4 V/µm, field induced second harmonic generation signal indicating the growth of oriented polar domains.\cite{nishikawa_fluid_2017}

An interesting open question relates to the origin of these polar nematic phases and what drives different materials to exhibit or not the intermediate antiferroelectric phase. We note here that for RM734 (Iso-N-N$_{\mathrm{F}}$ phase sequence) and DIO (Iso-N-N$_{\mathrm{S}}$-N$_{\mathrm{F}}$), the molecular dipole moment has a component perpendicular to the long molecular axis. A comparison of the relaxation processes observed in these two materials can be found in Fig. S3. Interestingly, for material $\mathbf{1}$ in an article by Manabe et al.,\cite{manabe_ferroelectric_2021} which shows a direct isotropic to N$_{\mathrm{F}}$ transition, the dipole is almost along the molecular long axis. In this context, and in combination with devoted models taking into account the intermediate phase, broadband dielectric spectroscopy could prove to be essential. An off-axis dipole moment implies that the nematic director and polarization do not need to be parallel. The two order parameters give rise to the in-phase and counter-phase fluctuations of the director and polarization as described by the CPM model,\cite{vaupotic_dielectric_2023} which can be rationalized as the collective precession around the cone and reorientations around the molecular short axis (Fig. \ref{fig:sketch}). Through the N-N$_{\mathrm{F}}$ phase transition of RM734, it has been shown that the splay mode is coupled with the magnitude of the polarization.\cite{PhysRevLett.124.037801} If the phase is modulated, as is the N$_{\mathrm{S}}$ phase, the twist mode, as depicted in Fig. S7, can also be seen dielectrically. For a periodicity of $p=18$ nm \cite{chen_smectic_2023} and diffusivities $K_2/\gamma$ in the range 10$^{-11}$ to 10$^{-12}$ m$^2$/s,  a relaxation frequency $f=(K_2/\gamma) (2\pi/p)^2$ of $0.1-1$ MHz is expected, where $K_2$ is the twist elastic constant and $\gamma$ is the rotational viscosity. However, experimental phase structural resolution together with a devoted theoretical model are needed for further understanding dielectrically accessible molecular correlations and their importance in the formation of the N$_{\mathrm{S}}$ phase.

\begin{figure}
\includegraphics[width=0.5\textwidth]{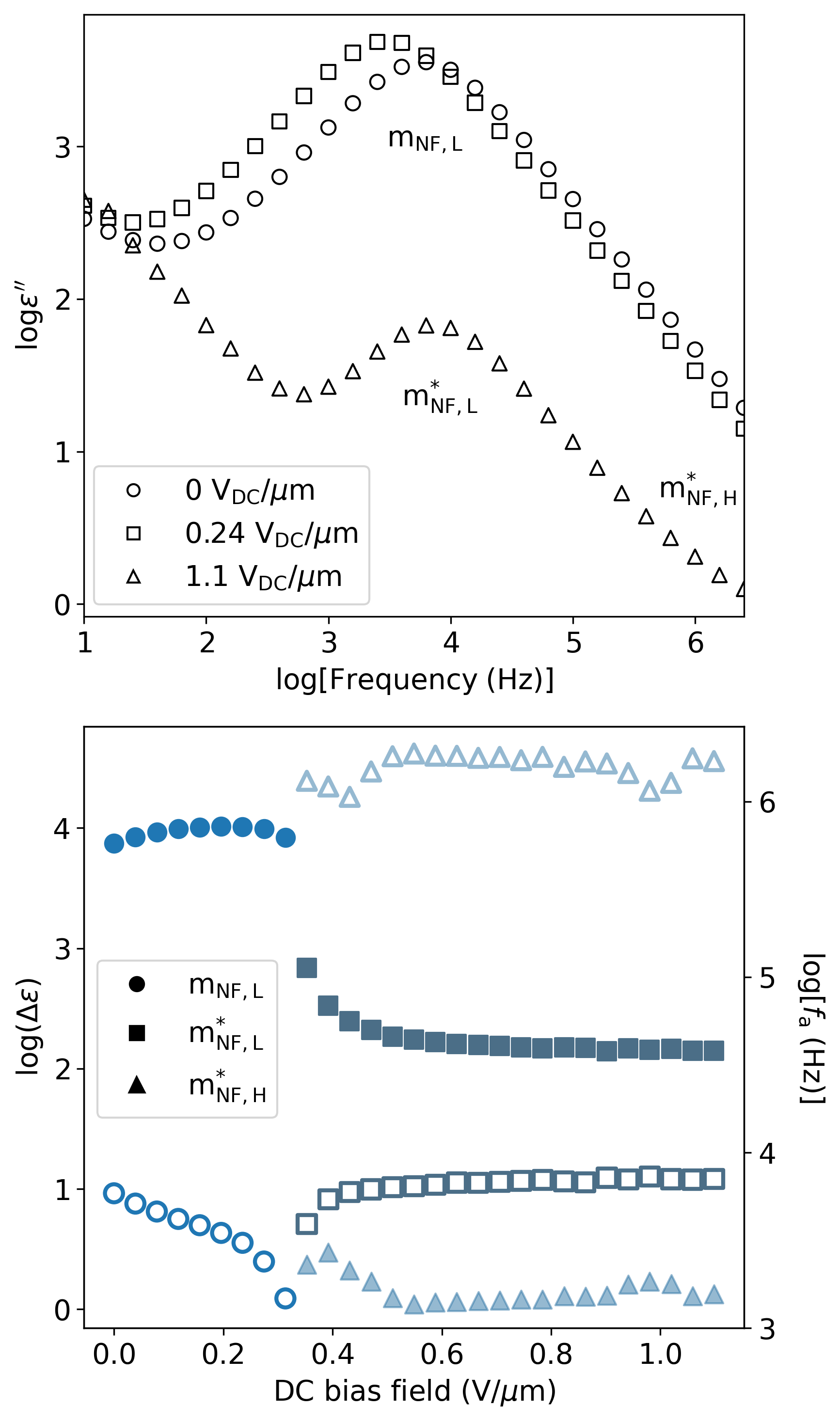}
\caption{\label{fig:fieldNF} Effect of a DC bias on the dielectric spectrum of the N$_{\mathrm{F}}$ phase at 65$^\circ$C in a 5.1 $\mu$m thick cell. Top: Dielectric loss spectrum at 0 V$_\mathrm{DC}$/$\mu$m, 0.24 V$_\mathrm{DC}$/$\mu$m and 1.1 V$_\mathrm{DC}$/$\mu$m. Bottom: Dielectric strengths ($\Delta\varepsilon$, solid symbols) and frequencies of maximum absorption ($f_a$, empty symbols) of the observed modes as a function of the external DC bias field.}
\end{figure}

In the N$_{\mathrm{F}}$ phase at 65$^\circ$C, upon progressive application of the DC bias field, the amplitude of m$_\mathrm{NF,L}$ slightly increases (from approximately $7\,000$ to $10\,000$) and then, above 0.3 V$_\mathrm{DC}$/$\mu$m, it gets partially suppressed and its relaxation frequency decreases (see Fig. \ref{fig:fieldNF}). The suppression of the mode reveals a second process of lower strength and slightly higher frequency (termed m$_\mathrm{NF,L}^*$). This behavior can be explained within the CPM model, in which the N$_{\mathrm{F}}$ phase spectrum would be dominated by the GM. However, when a DC bias field is applied, the dominant GM is suppressed and shifted to lower frequencies, and the SM, corresponding to fluctuations in the magnitude of the polarization, becomes visible at slightly higher frequencies. This was observed by Vaupotič et al. with a homologue of RM734.\cite{vaupotic_dielectric_2023} In our case we would identify m$_\mathrm{NF,L}$ as the dominant GM and m$_\mathrm{NF,L}^*$ as the SM. On the other hand, the reason for the slight increase of $\Delta \varepsilon$ at low DC bias fields (below 0.3 V$_\mathrm{DC}$/$\mu$m) might be the following: before the polarization vector and director are completely reoriented, the effect of the DC bias field, as suggested by Vaupotič et al., is to shrink the layer of LC material over which the polarization rotates to become parallel to the electric field.\cite{vaupotic_dielectric_2023} This would increase the capacitance of the dielectric layer and, thus, the strength of the mode should increase. This is precisely in line with the PCG mode proposed by Clark et al.\cite{clark_dielectric_2022} Lastly, returning to Fig. \ref{fig:fieldNF}, a high-frequency process of negligible DC bias field dependence appears (m$_\mathrm{NF,H}^*$). This could be the continuation of m$_\mathrm{I}$ in the N$_{\mathrm{F}}$ phase. The reorientation of the director under sufficiently strong bias fields was confirmed by POM (see Fig. S8).

\begin{figure}
\includegraphics[width=0.5\textwidth]{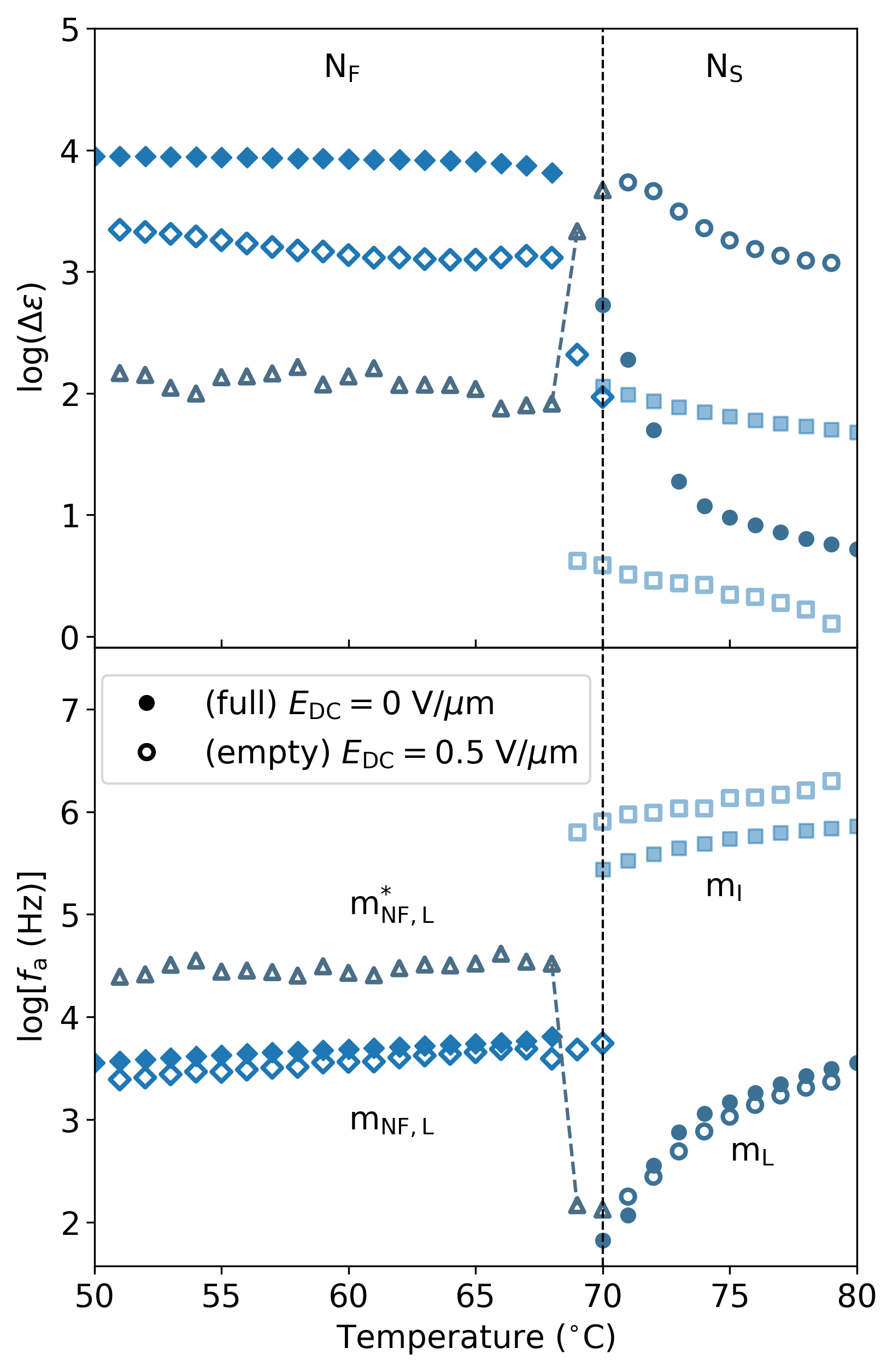}
\caption{\label{fig:DC-cool} Dielectric strengths ($\Delta\varepsilon$) and frequencies ($f_a$) of maximum absorption of the observed modes obtained from fits to the HN formula (Equation \ref{HN_eq}) with no bias field (solid symbols) and with a bias field of 0.5 V$_\mathrm{DC}$/$\mu$m (empty symbols) in a 5.1 $\mu$m thick cell. The dashed lines are guides for the eyes.}
\end{figure}

Finally, we studied the effect of an DC bias field applied while cooling on the dielectric spectra of the N$_{\mathrm{S}}$ and N$_{\mathrm{F}}$ phases. To minimize sample degradation due to long-time bias application, the cooling rate was set to 1$^\circ$C/min and we chose a 0.5 V$_\mathrm{DC}$/$\mu$m bias field to be well above the thresholds observed in Figs. \ref{fig:fieldNS} and \ref{fig:fieldNF}. Fig. \ref{fig:DC-cool} shows the temperature dependence of the relaxation processes frequency and strength in a 5.1 $\mu$m-thick cell, where the results with no DC bias are also plotted for comparison. In the N$_{\mathrm{S}}$ phase, m$_\mathrm{L}$ shows a much larger strength under an applied DC bias field, as expected from Fig. \ref{fig:fieldNS}, while its relaxation frequency is practically unaltered. In both cases, the SM behavior is present. This again suggests that m$_\mathrm{L}$ is indeed the high-temperature branch of the SM. Under a DC bias field, m$_\mathrm{I}$ is strongly suppressed and slightly shifted in frequency. A dedicated model for the N$_{\mathrm{S}}$ phase should be developed in order to fully understand the origin of this mode. In the N$_{\mathrm{F}}$ phase, two modes are present under an applied DC bias field. The lower frequency process is the suppressed m$_\mathrm{NF,L}$ (GM), which dominates the spectra at 0 V$_\mathrm{DC}$, while the higher frequency one is the previously mentioned m$_\mathrm{NF,L}^*$ (SM). It is interesting to note that cooling the sample under a DC bias field does not significantly shift the GM in frequency. Near the N$_{\mathrm{S}}$–N$_{\mathrm{F}}$ transition, m$_\mathrm{NF,L}^*$ shows soft mode behavior, as expected. In particular, it is interesting to note that the temperature dependence of the low-temperature branch of the SM is steeper than that of the high-temperature one. This is in fact predicted by the CPM model, although the high-temperature phase considered by the authors was the conventional N phase. \cite{vaupotic_dielectric_2023} Away from the N$_{\mathrm{S}}$–N$_{\mathrm{F}}$ transition, m$_\mathrm{NF,L}^*$ shows a negligible temperature dependence, just like the m$_\mathrm{NF,L}$.

\section{Conclusions}

In summary, we have shown that the ferroelectric nematic compound DIO exhibits several relaxation processes of great complexity, which could be traced back to the magnitude of polar correlations in all three nematic phases. By studying the dielectric spectra in a broad range of frequencies and temperatures, as well as under the application of external electric fields, we were able to assign the observed modes to different dipolar fluctuations. The highest frequency mode accessible in our experiments, denoted by m$_\mathrm{H}$, can be associated with the rotation of individual molecules around their long axis or $\beta$-process, and is present throughout the whole phase sequence. The lowest frequency mode in the N and N$_{\mathrm{S}}$ phases, called m$_\mathrm{L}$, is related to the flip-flop rotation of molecules or $\delta$-process, but it involves some degree of collective behavior, even in the paraelectric N phase. In the N$_{\mathrm{S}}$ phase, the dipolar correlations grow and are responsible for the amplitude fluctuations (SM). An intermediate-frequency process appears in this phase, called m$_\mathrm{I}$, which also exhibits SM behavior and might be associated with the twist mode. Finally, the GM and OM predicted by the CPM model are observed in the N$_{\mathrm{F}}$ phase. However, the GM is overlapped with the SM, which has a much lower amplitude, but can be made visible by applying a sufficiently strong DC bias field. The non-ideal nature of this GM, which appears at a non-zero frequency, is thought to be the origin of the large permittivity values and the thickness effect. In view of the present results, a realistic value for $\varepsilon'$ might be of the order $10^2$–$10^3$ due to the large dipole moment of DIO, as observed when suppressing polarization reorientation processes by applying a sufficiently strong DC bias field.

\section{Supplementary material}
The supplementary material includes additional dielectric spectroscopy data, fit examples and results, POM observations and a visual representation of the proposed twist mode.

\section{Acknowledgements}

A.E. and J.M.-P. acknowledge funding from the Basque Government Project IT1458-22. A.E. thanks the Department of Education of the Basque Government for a predoctoral fellowship (grant no. PRE\_2022\_1\_0104). A.M. and N.S. acknowledge funding from the Slovenian Research and Innovation Agency (grant no. P1-0192). M.H. and S.A. acknowledge the National Key Research and Development Program of China (no. 2022YFA1405000) and the Recruitment Program of Guangdong (no. 2016ZT06C322). We are grateful to Ander García Díez from  the Basque Center for Materials, Applications and Nanostructures (BCMaterials) for the preparation of the gold electrodes.

\section{Author Declaration}
\subsection{Conflict of Interest}
The authors have no conflicts to disclose.

\section{Data Availability}
The data that support the findings of this study are available within the article and its supplementary material.


%
%

%


\bibliography{REFERENCES}

\end{document}